\documentclass[prb,twocolumn,amsmath,amssymb,floatfix]{revtex4}
\usepackage{graphicx}
\usepackage{bm}

\newcommand{\be}{\begin{equation}}
\newcommand{\ee}{\end{equation}}
\newcommand{\beq}{\begin{eqnarray}}
\newcommand{\eeq}{\end{eqnarray}}

\begin{document}

\title{Evolution of the macroscopically entangled states in optical lattices}

\author{Anatoli Polkovnikov}
\email{anatoli.polkovnikov@yale.edu}
\homepage{http://pantheon.yale.edu/~asp28} \affiliation{Department
of Physics, Yale University, P.O. Box 208120, New Haven CT
06520-8120}

\date{\today}

\begin{abstract}
We consider dynamics of boson condensates in finite optical
lattices under a slow external perturbation which brings the
system to the unstable equilibrium. It is shown that quantum
fluctuations drive the condensate into the maximally entangled
state. We argue that the truncated Wigner approximation being a
natural generalization of the Gross-Pitaevskii classical equations
of motion is adequate to correctly describe the time evolution
including both collapse and revival of the condensate.
\end{abstract}

\maketitle
\section{Introduction}
\label{sec:1}

Recent advances in experimental realization of Bose-Einstein
condensates (BECs) in optical lattices~\cite{Kasevich, Bloch,
Bloch1} make this field particularly interesting for theoretical
analysis. One of the most striking features about these
condensates is the possibility to observe directly effects of
quantum fluctuations at zero temperature. For example, as was
predicted theoretically~\cite{Fisher} and shown
experimentally~\cite{Bloch}, the zero point motion can drive the
system from the superfluid to the Mott insulating state. The other
direct manifestation of the quantum effects was reported
in~Ref.[\onlinecite{Bloch1}], where it has been shown that bosons
can live in the superposition of number states even at the absence
of tunnelling. On the other hand, in the superfluid regime the
quantum fluctuations are suppressed and either classical
Gross-Pitaevskii (GP) or Bogoliubov approach is often adequate for
the description of both static and dynamic properties of the
condensates (see e.g. Refs.~[\onlinecite{Leggett, Rey}]). However,
there is an interesting possibility, wherein the system is
superfluid but neither of these approaches is good. Suppose that
the initially stable condensate is driven to the regime of
instability. This can be achieved either by applying a certain
phaseshift to the condensate with repulsive
interactions~\cite{Smerzi, Wu, Raghavan} or by switching the sign
of the interaction to the negative value  using Feshbach
resonance~\cite{Kanamoto, Kavoulakis} . The main difficulty arises
because near the instability all the fluctuations including
quantum exponentially diverge and cannot be treated as a small
perturbation. To be more specific, suppose that for time $t \leq
0$ the periodic system of condensates in a lattice was in a
superfluid ground state, i.e. the interaction was relatively weak.
Then a phase imprint, i.e. a certain phase difference between the
adjacent wells, was imposed. Experimentally it can be achieved by
e.g. applying a short (compared to a single tunneling time) pulse
of external field to the system. A case of special interest will
be when there is a relative $\pi$ phase shift between neighboring
wells~\cite{psg}. For the two wells with equal number of bosons
and relatively small interaction , this state is
metastable~\cite{Smerzi, psg} (this is also the case for even
number of wells and periodic boundary conditions). However, if the
interaction increases becomes larger than a critical value, this
equilibrium becomes unstable and the bosons spontaneously form a
``dipole'' state~\cite{psg,Vardi,Franzosi,Raghavan} in which most
of them occupy one of the two wells. Upon accounting for quantum
fluctuations in a system with a finite number of bosons, the state
obtained is a superposition of the two dipole states so that the
inversion symmetry is preserved. Clearly in the case of infinite
number of wells translational symmetry is always broken. For
example in Ref.~[\onlinecite{Sachdev2}] a similar instability but
for the case of a Mott insulator in a strong electric field was
shown to drive the system into a dipole state.

Related to this instability is a very interesting possibility of
forming a Schr\"odinger cat state~\cite{Kasevich1}. If the
interaction slowly increases in the $\pi$ state, then as we just
mentioned, at certain point the system becomes unstable.
Classically the bosons will remain in this unstable state forever
unless there is some noise present either dynamical or in the
initial conditions. As we will show below such a noise will drive
all the bosons into one spontaneously chosen well. However, apart
from classical fluctuations, which are always there but relatively
weak in the condensates, there is also a quantum zero point
motion, which comes from the uncertainty relation between the
number of bosons and their phase, so that the state where both are
defined is simply impossible. This quantum noise will also cause
the classical trajectories to move apart from the unstable
equilibrium. However, as we mentioned above, the quantum
fluctuations do not break translational invariance so the
resulting state must be macroscopically entangled. Let us give a
simple analogy with a ball laying on the top of the hill. Without
fluctuations it will remain there forever. However, because of the
uncertainty principle this ball will move down along different
classical paths. The quantum effects will be manifested only in
certain phase relations between these paths but will not affect
the motion itself. This analogy suggests that a good way to
describe these situations is to take into account fluctuations
yielding some probability distribution of the number and phase at
$t=0$ and evolve the fields according to the classical equations
of motion. In the literature this approach is known as the
truncated Wigner approximation (TWA)~\cite{Walls, Steel, Sinatra,
Sinatra1, Sinatra2, Lobo}. In the Appendix~\ref{twa} we will show
that this approach naturally arises in the path integral
derivation of the evolution equations, and in
Sec.~\ref{sec:quantum} we will numerically test the results on the
exactly solvable model of two condensates. We show how to go
beyond TWA in a separate publication~\cite{ap1}.

Let us also briefly mention a few other alternatives to generalize
the classical dynamics of the BECs existing in literature. One of
them is a generalization of GP equations by adding the interaction
of the superfluid component of the condensate with excited
bosons~\cite{Stoof1,Duine}. This method was successfully applied
for the description of the condensate evaporation after a sudden
change in the scattering length. It is hard to see though what is
the range of applicability of the resulting equations and how
crucial are the assumptions made about the dynamics of incoherent
excitations. Recently, there was developed another class of
methods based on exact stochastic dynamics~\cite{Steel,
Carusotto1, Carusotto2}. These ideas look very promising, but so
far the theory is applicable to a particular class of two-body
interactions and at least to author it is not completely clear how
to generalize it.

Throughout this paper we will explicitly consider a one
dimensional array of coupled condensates with. However, the
results are quite general and should not depend on dimensionality.

The standard Bose-Hubbard hamiltonian we are going to employ
reads:
\be
\mathcal{H}=\sum_j  -J(a_j^\dagger a_{j+1}+a_{j+1}^\dagger a_j )+
{U\over 2} a_j^\dagger a_j (a_j^\dagger a_j-1).
\label{hamilt}
\ee
Here $a_j$ is the canonical Bose annihilation operator on sites of
the optical lattice (wells) labelled by an integer $j$, $J$ is the
tunneling amplitude between neighboring lattice sites, $U>0$ is
the repulsive interaction energy between bosons in the same well.
Another important parameter in the problem is the mean number of
bosons per lattice site - $N$. Throughout this paper we consider
the case of large $N$, since it corresponds to the nearly
classical limit. A dimensionless measure of the strength of
interactions between the bosons is the coupling~\cite{psg}
\begin{equation}
\lambda \equiv \frac{UN}{J}.
\label{deflambda}
\end{equation}
Hereafter, except otherwise specified, we set $\hbar=1$ and $J=1$
so all the energies are given in units of $J$ and time has units
of $\hbar/J$. As we noted before~\cite{psg}, $\lambda\sim 1$
corresponds to the crossover from weakly to strongly interacting
superfluid and $\lambda\sim N^2$ corresponds to the quantum phase
transition to the Mott insulating phase. We will be interested
only in the superfluid regime and assume that $\lambda\ll N^2$.
Note that the hamiltonian (\ref{hamilt}) clearly has a time
reversal invariance. Besides the equations of motion:
\be
i{da_j\over dt}=-[{\cal H},a_i]
\label{eq3}
\ee
are invariant under the transformation: $t\to -t$, $\lambda\to
-\lambda$, and $a_j\to (-1)^j a_j$. So in the absence of energy
relaxation, which would break the time reversal symmetry, a $\pi$
phase-shift between neighboring sites is equivalent to the change
of the sign of the interaction from repulsive to attractive. This
equivalence is very useful for qualitative understanding of the
resulting instabilities.

\section{Semiclassical evolution of the phase modulated state}
\label{sec:2}

As we discussed in detail in Ref.~[\onlinecite{psg}], in the
classical description of the two coupled condensates the effective
motion of the number difference is equivalent to that of a
classical particle with a unit mass in the effective potential:
\be
U_{eff}=2 n^2(1 +\lambda \sqrt{1-n_0^2} \cos\theta_0)+\lambda^2
n^2\left({n^2\over 2}-n_0^2\right)=0,
\label{cat1}
\ee
where the ``coordinate'' $n$ represents the number difference
between the left and the right sites, $\theta$ is the relative
phase; $n_0\equiv n(t=0)$ and $\theta_0\equiv \theta(t=0)$ are the
initial conditions. In particular, if $n_0=0$ and $\theta_0=\pi$
then
\be
U_{eff}=2n^2(1-\lambda)+{\lambda^2n^4\over 2}.
\ee
Clearly the equilibrium $n=0$ becomes unstable if
$\lambda>\lambda_c=1$.

In a more general case of multiple wells a similar analysis can be
done. Because there are now many degrees of freedom a simple
representation of the motion using the effective potential becomes
impossible. Instead let us return to the GP version of the
equations~(\ref{eq3}):
\be
i{d \psi_j\over d t}=- (\psi_{j+1}+\psi_{j-1})+\lambda(t)\,
\psi_j^\star \psi_j^2,
\label{eq4}
\ee
where $\psi_j(t)$ is the semiclassical field corresponding to the
expectation value of the operator $a_j(t)$. Clearly the $\pi$
modulated state is a stationary solution for any interaction:
\be
\psi_{j}(t)=(-1)^j \mathrm e^{i\Theta (t)}.
\ee
The unimportant global phase $\Theta(t)$ is given by:
\be
\Theta(t)=-2t-\int_{0}^t \lambda(\tau)d\tau.
\ee
Similarly to the two well case this state becomes unstable when
the interaction  exceeds a certain critical
value~\cite{Smerzi,psg}:
\be
\lambda_c=2\sin^2 {\pi\over M},
\label{lambdac}
\ee
where $M$ is the number of the lattice sites in the array. The
origin of the instability becomes intuitively clear if we use a
dynamical symmetry mentioned above: $\lambda\to -\lambda$,
$\psi_j\to (-1)^j \psi_j^\star$. So the strong repulsive
interaction for the $\pi$ state is equivalent to the strong
attractive interaction for the symmetric state. The instability
for the attractive interaction is naturally
expected~\cite{Kanamoto,Kavoulakis}. To get more quantitative
results we consider a time evolution of fluctuations around the
$\pi$ state:
\be
\psi_j(t)=(-1)^j \,{\rm e}^{i \Theta(t)} (1+\xi_j(t)+i\eta_j(t)),
\label{eq8}
\ee
with $\xi_j$ and $\eta_j$ being the small real deviations from the
exact $\pi$ solution found above. Substituting (\ref{eq8}) into
(\ref{eq4}) and linearizing the resulting equations we obtain:
\beq
{d\xi_j\over dt}&=&\eta_{j+1}+\eta_{j-1}-2\eta_j\label{eq9}\\
-{d \eta_j\over
dt}&=&\xi_{j+1}+\xi_{j-1}-2\xi_j+2\lambda(t)\xi_j\label{eq10}.
\eeq
In the Fourier space this system is equivalent to a set of
decoupled second order differential equations:
\beq
{d^2\xi_q\over dt^2}=-16\sin^4{q\over
2}\xi_q+8\lambda(t)\sin^2{q\over 2}\xi_q,
\eeq
and
\be
\eta_q=-{1\over 4\sin^2{q\over 2}}{d\xi_q\over dt}.
\ee
In the case of adiabatically changing interaction we can write:
\be
\xi_q(t)=\xi_{0\,q}\mathrm e^{i\phi_q(t)}
\label{eq10a}
\ee
and neglect by the second derivative of $\phi_q$. Here $\xi_{0\,
q}$ is the initial amplitude of fluctuations. Substituting
(\ref{eq10a}) into (\ref{eq10}) we get:
\be
{d\phi_q\over dt}=\pm 4\sin^2{q\over 2}\sqrt{1-{\lambda(t)\over
2\sin^2{q\over 2}}}.
\label{eq11}
\ee
For simplicity we assume that the interaction increases in time
as:
\be
\lambda(t)={\lambda_0\over 1-\delta t},
\ee
where $\delta$ is the  parameter of adiabaticity and $\lambda_0$
is the initial interaction, which we assume to be small . This
type of $\lambda(t)$ dependence, in fact, corresponds to the
tunnelling exponentially decreasing in time $(J(t)=J_0\mathrm
e^{-\delta t })$ with $t\to t/J(t)$. It is straightforward to
solve (\ref{eq11}) explicitly and the result is:
\beq
&&\phi_q(\lambda)=\pm {2\sin^2{q\over 2}\over
\delta}\Biggl[{\sqrt{4-{2\lambda_0\over \sin^2{q\over 2}}}\over
\lambda_0}-{\sqrt{4-{2\lambda\over \sin^2{q\over 2}}}\over
\lambda}\nonumber\\
&&+\ln{1\!+\!\sqrt{1-{\lambda\over 2\sin^2{q\over 2}}}\over
1-\sqrt{1\!-\!{\lambda\over 2\sin^2{q\over 2}}}}-
\ln{1\!+\!\sqrt{1-{\lambda_0\over 2\sin^2{q\over 2}}}\over
1\!-\!\sqrt{1-{\lambda_0\over 2\sin^2{q\over 2}}}}\Biggr].
\eeq
Assuming that $\lambda_0<2\sin^2{q/2}$ we see that in the limit
$\lambda\to\infty$ the imaginary part of phase $\phi_q$ goes to:
\be
\Im\phi_q(\infty)=\pm {2\pi\over \delta}\sin^2{q\over 2}.
\ee
So if $\delta$ is large enough then the instability cannot develop
in time and the phase remains essentially real. In the opposite
limit the fluctuations become large and we have to study the
nonlinear regime of GP equations. More specifically the relation
\be
|\xi_{0\,q}|\mathrm e^{\Im\phi_q(\infty)}=1
\ee
defines the boundary between the regimes of small and large
fluctuations. Using the estimate of $\xi_{0\,q}$ (see the next
section for the details)
\be
\xi_{0\,q}\sim {1\over \sqrt{N}}
\ee
we derive that the instability for a given momentum mode will
evolve into the nonlinear regime given that
\be
\delta\leq {2\pi\sin^2 {q\over 2}\over \ln N}.
\label{delta}
\ee
This tells us that the interaction should indeed change slowly in
time (at least near the onset of instability) and so justifies the
adiabatic limit we used. The lowest energy mode corresponds to the
momentum $q=2\pi/M$ so the lower boundary for $\delta$ becomes:
\be
\delta_1={2\pi\over \ln N}\sin^2{\pi\over M},
\ee
and the upper boundary is
\be
\delta_2={2\pi\over \ln N}.
\ee
If $\delta<\delta_1$ then all the modes have enough time to get
into the nonlinear regime. On the contrary if $\delta>\delta_2$
then the fluctuations around GP state will remain small. In the
intermediate regime $\delta_1<\delta<\delta_2$ some of the
momentum modes will exhibit small fluctuations and some will
become strongly enhanced.

\section{Quantum fluctuations}
\label{sec:quantum}

\subsection{Truncated Wigner approximation}

Here we will try to examine the role of quantum fluctuations.
Before doing actual calculations, let us give some qualitative
discussion. As we mentioned before we are interested in the
regime, where $N$ is large and interactions are relatively weak
$\lambda\ll N^2$ so that the system is far from the Mott
insulating transition and the quantum fluctuations are
intrinsically small. This means that normally it is possible to
use GP approach or at most the Bogoliubov extension. However, this
is not the case for our problem. Indeed, near the classical
instability the starting point of unstable equilibrium for the
Bogoliubov expansion of the uniform condensate becomes bad. The
other way to describe this is to note, that the Bogoluibov
equations are nothing but the quantized version of linearized
equations (\ref{eq9},\,\ref{eq10}), which can predict the onset of
the instability but fail to describe the nonlinear regime. On the
other hand, we can anticipate that the quantum fluctuations will
remain weak until we cross the instability point. After that they
will force the system to evolve into the superposition of unstable
classical trajectories and become unimportant again, when those
trajectories will be relatively far from each other.

These ideas known as a truncated Wigner approximation~\cite{Walls}
have been recently applied for the description of
BECs~\cite{Steel,Lobo,Sinatra,Sinatra1,Sinatra2}.  The usual
method of deriving this scheme is based on the cubic Fokker-Planck
equations of motion for the density matrix written in the Wigner
representation. The main difficulty with the these Fokker-Planck
equations is that the third order derivative terms, which are
responsible for the corrections to GP trajectories, makes them
practically intractable~\cite{Steel}. Another entirely different
Keldysh technique can be also used to attack non-equilibrium
problems~\cite{Kamenev}. Classical equations of motion appear
there as the saddle point of the action. However any attempt to
include quantum fluctuations immediately results in a path
integral with a self-interacting classical field~\cite{Kamenev},
which is very complicated. In the Appendix~\ref{twa} we will show
how the TWA naturally arises from the path integral formulation of
the dynamics and emphasize the key difference between the present
derivation and that of the conventional Keldysh technique.
Moreover, in~Ref.[\onlinecite{ap1}] we show that it is
straightforward to go beyond TWA perturbatively including quantum
effect on the classical trajectories themselves. We will also
argue there that the TWA gives exact short-time asymptotical
behavior of the evolution of {\em any} system, the time when it
breaks down, however, depends on the details of the particular
process (see Sec.~\ref{subsec:two}).

The whole idea of the TWA is that the expectation value of any
given operator $\Omega$ at time $t$ is equal to the corresponding
classical observable $\Omega_{cl}(t)$ evaluated according to
standard GP equations and averaged over an ensemble of initial
conditions distributed according to the Wigner transform of the
initial density matrix (see Appendix~\ref{twa} for the details of
the derivation):
\be
\Omega(t)=\int d\psi_0^\star d\psi_0 p(\psi_0,\psi_0^\star)
\Omega_{cl}(\psi(t),\psi^\star(t),t),
\label{eq25}
\ee
where $p$ is defined as:
\beq
&&p(\psi_0,\psi_0^\star)=\int d\eta_0^\star d\eta_0 \langle
\psi_0-{\eta_0\over 2}|\rho_0 | \psi_0+{\eta_0\over
2}\rangle\nonumber\\
&&~~~~~~~~~~~~~~~~~\times \mathrm e^{-|\psi_0|^2-{1\over
4}|\eta_0|^2}\,\mathrm e^{{1\over
2}(\eta_0^\star\psi_0-\eta_0\psi_0^\star)}.
\label{eq26a}
\eeq
In the equation above $|\psi_0\pm\eta_0/2|$ denote coherent
states. We use the following measure
\be
d\psi_0 d\psi_0^\star\equiv \prod_\alpha {d\Re\psi_{0\,\alpha}\,
d\Im\psi_{0\,\alpha}\over \pi}
\ee
with the product taken over continuous or discrete spatial
indices, which we suppressed in (\ref{eq25}) and (\ref{eq26a}) to
shorten the notations. The interpretation of
$p(\psi_0,\psi_0^\star)$ as a probability is not very precise,
because the Wigner transform does not have to be positive. To get
the function $\Omega_{cl}(\psi,\psi^\star,t)$ we need to rewrite
the quantum operator $\Omega$ in the fully symmetrized form and
substitute field operators $a$ and $a^\dagger$ by their classical
counterparts $\psi$ and $\psi^\star$. In particular, the relation
between $\Omega_{cl}$ and a more familiar version of the classical
counterpart of the normal ordered operator $\Omega$ usually
appearing in the functional integrals is:
\be
\Omega_{cl}=\langle \Omega\left(\psi^\star+{\eta^\star/
2},\psi-{\eta/ 2}\right)\rangle,
\label{eq26c}
\ee
where the average is taken over $\eta$ with the weight
$\exp(-|\eta|^2/2)$.

Let us now give general comments on validity of (\ref{eq25}). If
the Hamiltonian is non-interacting, then this expression is exact.
So TWA includes the Bogoliubov approximation and goes beyond. To
recover the latter we just need to linearize the classical GP
equations of motion while evaluating $\psi(t)$. This statement is
not surprising since the noninteracting evolution is always
identical to classical~\cite{Caldeira, Gardiner, Sinatra2}. If
there are nonlinear interactions, then in general, there will be
corrections to the equations of motion themselves. We consider
them in~Ref.[\onlinecite{ap1}]. Let us only note here that for the
two coupled condensates we showed in~Ref.[\onlinecite{psg}] (see
also Sec.~\ref{subsec:two}) that the time where GP breaks down in
the worst possible scenario with the least classical initial state
having completely undefined phase is equal to $t_c\approx
N/\lambda=J/U$. We expect that the scaling with $N$ is
generic\footnote{To get the unique classical limit at $N\to\infty$
we should keep $\lambda$, not $U$, to be independent of $N$.} and
therefore (\ref{eq25}) should be valid at least for the times
shorter than $t_c$. In next section we will see that if there are
no sudden perturbations, so that a small fraction of quantum
levels is populated then the time scale of validity of TWA becomes
much longer.

\subsection{Two coupled condensates. Comparison with exact solution.}
\label{subsec:two}

The main purpose of this section is to test the truncated Wigner
approximation on a simple example of two coupled condensates,
where it is straightforward to obtain the exact solution. The
two-well version of the hamiltonian (\ref{hamilt}) reads:
\be
\mathcal H(t)=- J a^\dagger_\alpha \tau_x^{\alpha\beta}
a_\beta+{U(t)\over 2}\,a_\alpha^\dagger a_\alpha(a_\alpha^\dagger
a_\alpha-1),
\label{ham1}
\ee
where $\alpha,\beta=L,R$ denote the right or the left well
respectively, $\tau_{a}^{\alpha\beta}$, $a=x,y,z$ are the Pauli
matrices. As usually we imply implicit summation over repeated
indices. Because of total number conservation (\ref{ham1}) is
equivalent to a more familiar version
\be
\tilde{\mathcal H}(t)=-J a^\dagger_\alpha \tau_x^{\alpha\beta}
a_\beta+{U(t)\over
4}\,(a_\alpha^\dagger\tau_z^{\alpha,\beta}a_\beta)^2,
\label{ham2}
\ee
A convenient choice of the observable is
\be
\Omega={1\over
N^2}(a_\alpha^\dagger\tau_z^{\alpha\beta}a_\beta)^2={1\over N^2}
:(a_\alpha^\dagger\tau_z^{\alpha\beta}a_\beta)^2:+{2\over N},
\label{eq29}
\ee
where semicolons denote the normal order. The operator $\Omega$ is
nothing but the variance of the relative number distribution. Let
us consider several examples of evolution: (i) the initial state
is symmetric and the interaction increases with time, (ii) the
initial state is antisymmetric and the interaction increases with
time, and (iii) the initial state is the Fock state and the
interaction does not change in time. The situation (ii) is
directly relevant to the ``cat'' state dynamics we consider in
this paper, but we also look to the other possibilities to check
the validity of this approach in a more general case.

The classical function $\Omega_{cl}(\psi,\psi^\star,t)$ can be
either found from the normal-ordered form of the operator $\Omega$
according to (\ref{eq26c}) or by direct symmetrization of the
latter. In our particular case it reads:
\beq
\Omega_{cl}(\psi^\star,\psi)&=&(\psi_\alpha^\star
\tau_z^{\alpha\beta}\psi_\beta)^2+{2\over N} -
{2\over N}-{1\over 8N^2} \nonumber\\
&=&(\psi_\alpha^\star\tau_z^{\alpha\beta}\psi_\beta)^2-{1\over
8N^2}.
\label{eq30}
\eeq
The final step is to find the probability function
$p(\psi_0,\psi_0^\star)$ according to~(\ref{eq26a}). This will
depend on the details of the state $|O\rangle$, therefore we have
to study different initial configurations explicitly.

\subsubsection{Symmetric or antisymmetric initial state}

\begin{figure}[ht]
\centering
\includegraphics[width=8.3cm]{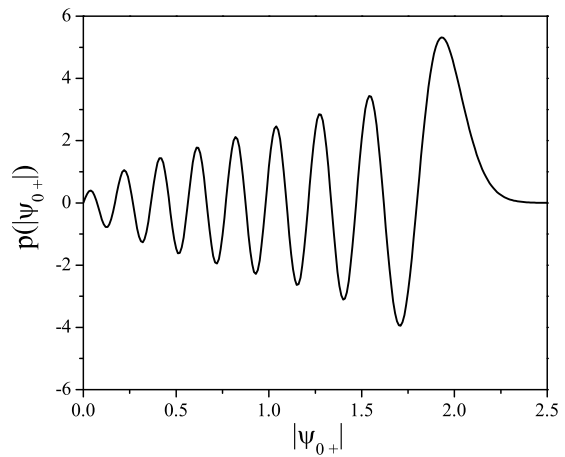}
\caption{Distribution of the GP initial conditions versus
$|\psi_{0\,+}|$ for the symmetric state with eight noninteracting
bosons per well.}
\label{cat:fig1}
\end{figure}

Suppose that at $t=0$ the interaction was negligible. Then the
products of symmetric and antisymmetric wavefunctions give the
ground and the most excited stationary states, respectively. They
can be also represented as a superposition of products of the two
coherent states with equal or shifted by $\pi$ phases:
\be
|0\rangle=(4\pi N)^{1/4}\mathrm e^{-N} \int {d\theta\over 2\pi}
\mathrm e^{-2i\theta N}|\sqrt{N}\mathrm e^{i\theta}\rangle_L
|\sqrt{N}\mathrm e^{i\theta+i\pi\sigma}\rangle_R,
\label{eq31}
\ee
where $\sigma=0,1$ for the symmetric or antisymmetric state,
respectively. The integral over the global phase $\theta$ ensures
the particle number conservation. Before proceeding with further
analysis let us look into a simpler example of just a product of
the two coherent states, where the global phase symmetry is broken
and $\theta$ takes some particular value. Then after
straightforward calculation one can show that:
\be
p(\psi_0,\psi_0^\star)=4\mathrm e^{-2\sum_\alpha
|\psi_{0\,\alpha}-\sqrt{N}\mathrm e^{i\pi\sigma}|^2}.
\label{eq32}
\ee
We see that in this case the probability distribution of $\psi_0$
is just a gaussian centered near the classical value with the
relative variance of fluctuations of the order of $1/\sqrt{N}$.
This is completely reasonable and we indeed recover GP picture
having a single initial state in the limit $N\to\infty$. Now let
us look closer to the wave function (\ref{eq31}). After a simple
calculation the final expression for the probability $p$ reads:
\be
p(\psi_0,\psi_0^\star)=4{\rm
e}^{-|\psi_{0\,+}|^2-|\psi_{0\,-}|^2}L_{2N}(2|\psi_{0\, +}|^2),
\label{eq34}
\ee
where $\psi_{0\,\pm}=\psi_{0\,L}\pm\psi_{0\,R}$ in the symmetric
state, and we should interchange $\psi_{0\,+}$ and $\psi_{0\,-}$
for the $\pi$ state, $L_{2N}(x)$ is the Laguerre's polynomial.
This expression is a gaussian in terms of $|\psi_{0\,-}|$, however
it has a nonlocal behavior as a function of $|\psi_{0\,+}|$.
Moreover $p(\psi_0,\psi_0^\star)$ is not positively defined. This
leads to interesting consequences. For example, while the average
of $|\psi_{0\,+}|^2$, computed with help of (\ref{eq34}) gives the
expected classical result, the variance of $|\psi_{0\,+}|^2$ is
{\em negative}. In Fig.~\ref{cat:fig1} we plot the normalized
function $|\psi_{0\,+}|\, p(|\psi_{0\,+}|)$ for the situation with
8 bosons per well. The extra factor of $|\psi_{0\,+}|$ comes from
the integral measure:
\be
\int_0^\infty |\psi_{0\,+}|p(|\psi_{0\,+}|) d|\psi_{0\,+}|=1,
\ee
For convenience we rescaled the fields $\psi_0\to \sqrt{N}\psi$ so
that the classical expectation value of $|\psi_{0\,+}|$ is 2.
In the limit $N\to\infty$ we again recover the classical result
(for the rescaled fields) $\psi_{0\,L,R}=1$ or $|\psi_{0\,+}|=2,\;
|\psi_{0\,-}|=0$, but in a peculiar way. The contributions from
$|\psi_{0\,+}|<2$ will cancel each other because of fast
oscillations of the probability $p$ and only the small interval
around $|\psi_{0\,+}|=2$ will give the contribution to the final
result. We might think, that if the observable is a smooth
function of the initial parameters, then the details of the
distribution $p(\psi_0,\psi_0^\star)$ are not important and we can
substitute it by some gaussian function with appropriate mean and
variance. However, as we pointed out before the variance given by
the distribution (\ref{eq34}) is negative, so at least this is not
very straightforward to do.

\begin{figure}
\centering
\includegraphics[width=8.3cm]{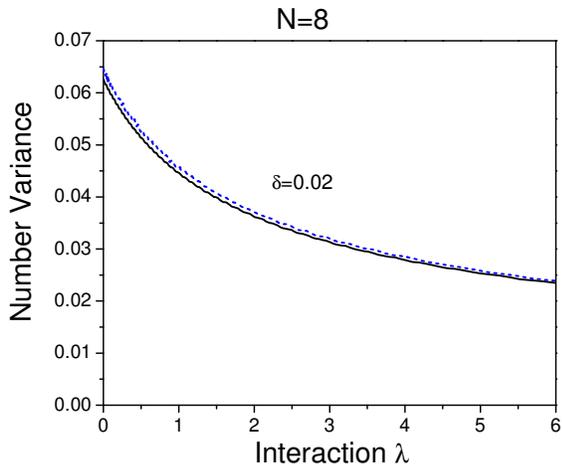}
\caption{Dependence of the number variance on the interaction
changing with time according to (\ref{eq36}) for initial symmetric
state and 8 bosons per well. Dashed and solid lines show
semiclassical and exact solutions, respectively. The lower graphs
correspond to a faster turning on the interaction. }
\label{cat:fig2a}
\end{figure}

\begin{figure}
\centering
\includegraphics[width=8.3cm]{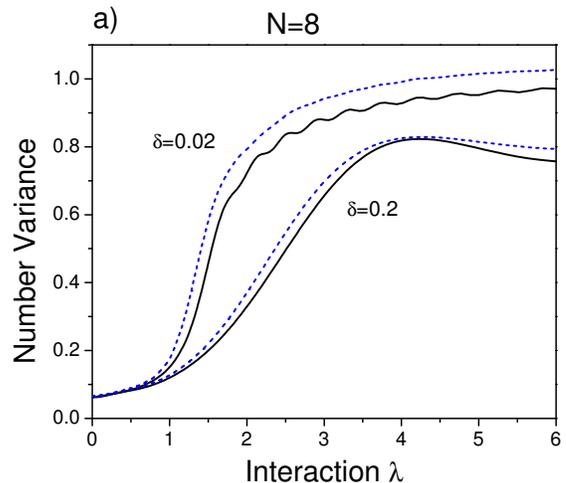}
\includegraphics[width=8.3cm]{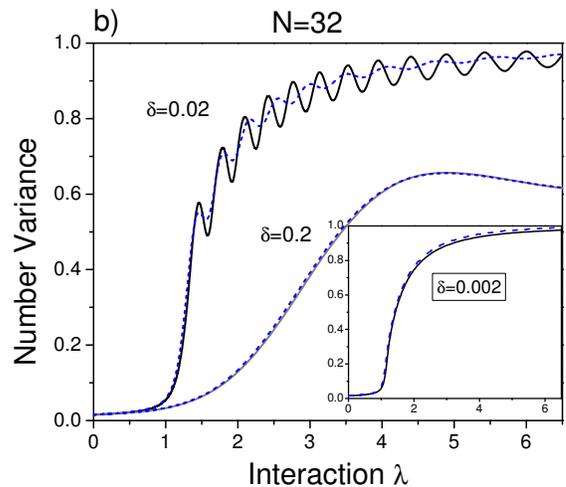}
\caption{Same as in Fig.~\ref{cat:fig2a} but for initial
antisymmetric state. The graphs (a) and (b) correspond to 8 and 32
bosons per well, respectively . The inset on the graph (b)
corresponds to the slowest evolution.}
\label{cat:fig2b}
\end{figure}

Now let us assume that the interaction increases with time
according to:
\be
\lambda(t)={\tanh(\delta t)\over 1-\delta t},
\label{eq36}
\ee
where $\delta\ll 1$ is the adiabatic parameter. This dependence is
somewhat different from what we used in the previous section. But
the resulting instability is still there, and besides the main
purpose of this section is to test our approximation scheme rather
then to do some particular calculations. The resulting graphs for
both symmetric and antisymmetric initial states are plotted in
Figs.~\ref{cat:fig2a} and ~\ref{cat:fig2b}. Note that even for the
eight particles per well the agreement between the exact and the
TWA solutions is remarkable. For the thirty two particles there is
a small discrepancy for the intermediate value $\delta$.
Apparently the semiclassical curve does not capture the small
oscillations very well. But note that both in the limit of large
and small $\delta$ the oscillations disappear and the agreement
becomes perfect.

Notice that the steady state for the initial antisymmetric
conditions is exactly the maximally entangled ``Schr\"odinger
cat'' state, where all the bosons occupy either left or right
well:
\be
|\Psi_f\rangle={1\over \sqrt{2}}
(|LLL\dots\rangle+|RRR\dots\rangle).
\ee
The ultimate reason for this is that as we mentioned above the
$\pi$ shifted state is at classical equilibrium for any
interactions  $\lambda$. This equilibrium though becomes unstable
for $\lambda>\lambda_c$. So any fluctuation will cause a classical
trajectory to end up either in the left or in the right well and
the quantum zero point motion gives us these fluctuations.
However, quantum mechanically we do not break the left-right
symmetry, because it is the property of the full hamiltonian. The
only way to reconcile these two results is to have the final
configuration in the coherent superposition of the left and the
right states. This statement can be also verified numerically.

\subsubsection{Initial number state}
\label{sec:num}

\begin{figure}[ht]
\centering
\includegraphics[width=8.3cm]{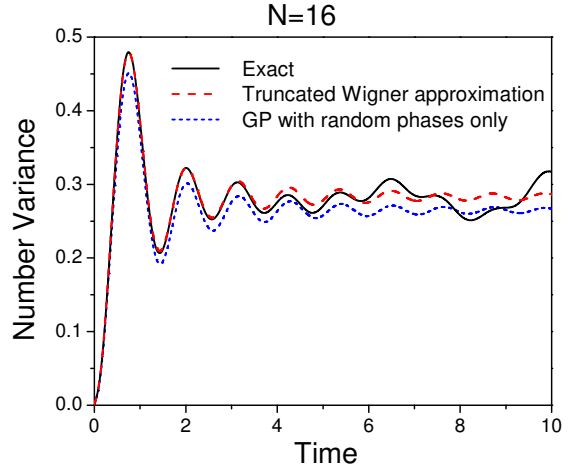}
\caption{Time dependence of the number variance for the initial
number state and 16 bosons per well. The interaction strength is
$\lambda=1$. Note that here and elsewhere in this paper time is
measured in units of $\hbar/J$.}
\label{cat:fig3}
\end{figure}

Here we revisit our results derived earlier~\cite{psg} assuming
the two condensates are initially uncoupled and their wavefunction
is just a product of the two number (Fock) states. Then at $t=0$
the tunnelling is suddenly turned on and the number variance
starts to experience some oscillatory behavior~\cite{psg}.
Repeating the same analysis as in the previous section we find:
\beq
&&\langle
(a_\alpha^\dagger(t)\tau_z^{\alpha,\beta}a_\beta(t))^2\rangle=\int_0^\infty
\int_0^\infty \int_0^{2\pi} dn_L\,dn_R\,  d\theta \nonumber\\
&&~~~~~~~~~p_{num}(n_L)p_{num}(n_R)
(\psi_\alpha^\star(t)\tau_z^{\alpha,\beta}\psi_\beta(t))^2,\label{aaa}
\eeq
where $n_{L,R}=|\psi_{L,R}(t=0)|^2$, $\theta$ is the initial phase
difference between $\psi_L$ and $\psi_R$. So in the Fock state the
phases in the two wells are indeed uncorrelated as we argued
in~Ref.[\onlinecite{psg}]. However the number of bosons is
distributed according to $p_{num}(n)$ given below and not fixed at
$n=N$ as we might naively think. In (\ref{aaa}) we ignored an
additive $1/8N^2$ correction (see (\ref{eq30})). The probability
of having initial occupation $n$ in either well
is~\cite{Gardiner}:
\be
p_{num}(n)=2\mathrm e^{-2 n}L_N(4n),
\label{eq38}
\ee
where as before $L_N(x)$ stands for Laguerre's polynomial of the
order $N$. The function $p_{num}$ is very similar to its
counterpart defined in (\ref{eq34}) in the sense that it has also
an oscillatory behavior for $n<N$ and exponentially decays for
$n>N$. In~Ref.[\onlinecite{psg}] we showed that the simple GP
picture (i) gives a multiplicative error ($1+2/N$) in the number
variance even in the noninteracting limit, and (ii) it is valid
for the finite amount of time shorter than some characteristic
scale determined by interactions: $t<t_c\approx J/U=N/\lambda$.
For longer times the GP result starts to deviate strongly from the
exact solution due to recurrence occurring in a quantum system. We
might guess that the agreement between the semiclassical and the
quantum results can be improved upon including quantum
fluctuations at initial time according to (\ref{eq38}). This is
indeed the case for $t<t_c$, i.e. the discrepancy in the prefactor
completely disappears. However, as we can see from
Fig.~\ref{cat:fig3}, these fluctuations do not affect the time
$t_c$ itself so that the correct result can be recovered only if
we also include quantum scattering, or in other words deviations
of the trajectories from the classical ones. Finally we would like
to note that the initial number state is the worst possible from
the classical point of view, because the phase is completely
undefined there. So we expect that in general $t_c$ gives the
lower boundary of the applicability of TWA.

\section{Numerical Results for Multiple Wells}
\label{sec:four}

Having established the general framework and checked its validity
let us move on to the main subject of the paper. First, following
the analysis given in Sec. \ref{sec:2}, we will study the temporal
behavior in a periodic array of wells being initially in the $\pi$
state. Then we will consider a case of a harmonic trapping
potential.

\subsubsection{Periodic Array}

The straightforward generalization of (\ref{eq31}) for the $\pi$
state in a periodic chain of $M$ coupled condensates with $N$
bosons per well (we assume $M$ to be even) is:
\be
|0\rangle=(4\pi N M)^{1\over 4}\mathrm e^{-{NM\over 2}} \int
{d\theta\over 2\pi} \mathrm e^{-i\theta NM}\prod_{j=1}^M
|\sqrt{N}\mathrm e^{i\theta+i\pi j}\rangle_j ,
\label{eq39}
\ee
where $|\dots \rangle_j$ stands for the coherent state in the
$j$-th well. This is an eigenstate of the noninteracting
hamiltonian and apart from the global phase $\theta$, which
conserves the total number of bosons, it is just a product of
coherent states with alternating phases. Ignoring the integral
over $\theta$, results in a gaussian probability distribution of
the initial state (compare with (\ref{eq32})):
\be
p(\psi_0,\psi_0^\star)=2^M\prod_{j=1}^M\mathrm
e^{-2|\psi_{0\,j}-\sqrt{N}\mathrm e^{i\pi j}|^2}.
\label{eq40}
\ee
while the correct result for (\ref{eq39}) reads:
\be
p(\psi_0,\psi_0^\star)=2^M\mathrm e^{-2 M\sum_k
|\hat\psi_{0\,k}|^2} L_{NM}(4M|\hat\psi_{0\,0}|^2),
\label{eq41}
\ee
where $\hat\psi_{0\,k}$ stands for the discrete Fourier transform
of $\psi_{0\,j}$:
\be
\hat \psi_{0\,k}={1\over M}\sum_j \psi_{0\,j} \mathrm e^{-ikj}
\ee
Clearly as $M\to\infty$ the difference between (\ref{eq40}) and
(\ref{eq41}) should vanish. Next let us define the operator
$\Omega$, which would be a good measure of the instability:
\be \Omega={1\over N^2 M(M-1)}\sum_j
(a_j^\dagger a_j-N)^2.
\ee
This is just a normalized sum of number variances over the
different wells. We have chosen the prefactor so that $\langle
\Omega\rangle\leq 1$, with $1$ corresponding to the state with all
the bosons located in any single well. It is easy to verify that
the classical counterpart of $\Omega$ is:
\beq
\Omega_{cl}(\psi,\psi^\star)&=&{1\over N^2 M(M-1)}\sum_j
\left(\psi_j^\star\psi_j-N-{1\over 2}\right)^2\nonumber\\
&-&{1\over 4N^2(M-1)}.
\eeq
It is reasonable to expect that as the number of wells increases
the global phase becomes less and less important and therefore
(\ref{eq40}) becomes more and more accurate.

\begin{figure}[ht]
\centering
\includegraphics[width=8.3cm]{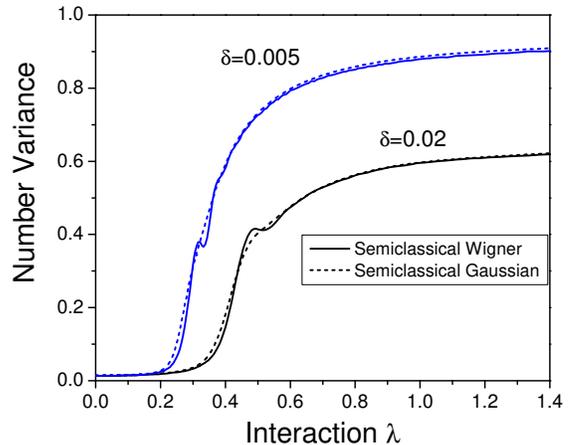}
\caption{Number variance as a function of interaction for the case
of ten wells with eight bosons per well. The solid and the dashed
lines correspond to the distributions given by (\ref{eq41}) and
(\ref{eq40}), respectively. The upper curves correspond to a
slower increase in interaction.}
\label{cat:fig4}
\end{figure}

Next we consider several specific examples. First let us take the
interaction to be monotonically increasing in time according
to~(\ref{eq36}). In Fig.~(\ref{cat:fig4}) we plot the resulting
evolution of the state for the case of ten wells with eight bosons
per well. The solid and the dashed lines correspond to the
probability distributions given by (\ref{eq41}) and (\ref{eq40}),
respectively . Clearly there is no significant difference between
them. Note that the upper curves corresponding to smaller
$\delta$, i.e. to the adiabatic limit, saturate at $\Omega\approx
1$, which shows that {\em all} the bosons appear in a single well,
i.e. the resulting steady state $|F\rangle$ is:
\be
|F\rangle={1\over \sqrt{M}}(|111\dots\rangle +
|222\dots\rangle+\dots |MMM\dots\rangle).
\ee
The whole procedure of driving the system into the maximally
entangled state described here is conceptually very similar to
that recently suggested in~Ref.[\onlinecite{Dorner}], where the
tunneling was assumed to decrease by the spatial drag of the
double-well condensate through a beam splitter. The important
difference however, is that here we are not limited by the
double-well system and can consider larger arrays, so that our
entangled ``cat'' occupies more than two macroscopic states.

\begin{figure}[ht]
\centering
\includegraphics[width=8.3cm]{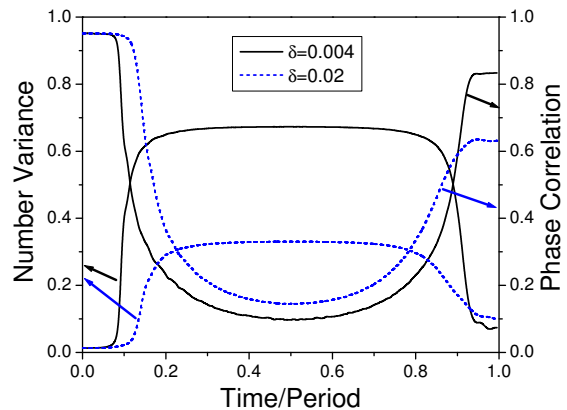}
\caption{Number variance and the phase correlation as a function
of time divided by the period of interaction ($T=1/\delta$) for
ten wells and eight bosons per well. The interaction changes with
time according to~(\ref{addt}). For smaller $\delta$ the phase
restoration is almost complete, which proves the coherence of the
dynamics.}
\label{cat:fig4a}
\end{figure}

There is an important issue, which was completely obscured in the
preceding analysis. Indeed, studying the number variance alone, it
is impossible to distinguish the ``cat'' state from the collapsed
condensate. While the collapse is often very well reproduced using
GP equations, it is much harder to describe the recovery within
this framework. To examine this issue let us consider the
interaction, which is periodic in time:
\be
\lambda(t)=\lambda_0\sin^2(\pi\delta t),
\label{addt}
\ee
where the parameter $\delta$ as in (\ref{eq36}) determines the
adiabaticity of the process. If $\delta$ is small then we expect
complete restoration of the initial state after one period of
oscillation $T=1/\delta$. With decreasing period we gradually
loose adiabatic limit and the evolution of the system is no longer
expected to be periodic. Fig.~\ref{cat:fig4a} summarizes this
discussion and the graphs are in perfect agreement with our
expectations. The phase correlation in this figure is defined in a
usual way as
\begin{displaymath}
-{1\over 2NM}\sum_j \langle a_j^\dagger a_{j+1}\rangle +c.c
\end{displaymath}
and the ``$-$'' sign is inserted for the sake of convenience
because we start from the $\pi$-state.

\begin{figure}[ht]
\centering
\includegraphics[width=8.3cm]{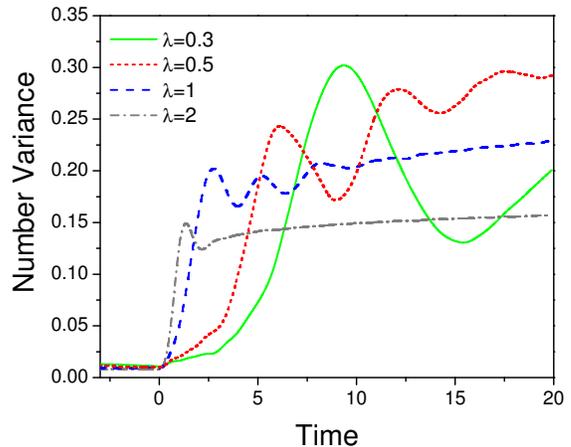}
\caption{Number variance as a function of time for ten wells and eight
bosons per well. The interaction remains constant in time in this
example but at $t=0$ a sudden $\pi$-phaseshift is applied to the
system.}
\label{cat:fig5a}
\end{figure}

So far we considered examples, where the initial state was a
noninteracting $\pi$-phaseshifted condensate and then the
interaction was slowly ramped up (or equivalently tunnelling was
ramped down). This rather hypothetical situation is certainly
suitable for the theoretical analysis, but hardly applicable to
any real experimental realization, where the interactions between
the bosons is always present. So the initial wavefunction is not a
simple product of coherent states but a more complicated object.
One way to proceed with this issue is to write down an approximate
wavefunction, which can be obtained using either Bogoliubov
hamiltonian (valid for $\lambda\ll N^2$) or variational methods,
and then find its Wigner transform according to~(\ref{eq26a}).
This would be a tractable but lengthy calculation. Instead we can
do a simple trick, demonstrating the advantage of the present
approach. Namely, we can start from the simple noninteracting
ground state, then adiabatically increase the interaction strength
to the desired value, and finally apply the $\pi$-phaseshift and
follow the subsequent evolution. In this scheme there is no need
to do any additional analytic calculations, both the ``fake''
evolution to the interacting ground state and the subsequent real
dynamics are described within the same scheme. Moreover the
computational time does not increase much because of the extra
``fake'' evolution to the true ground state. Indeed, the classical
motion is stable before the $\pi$-phaseshift is applied so that GP
equations can be efficiently integrated. To be more specific,
assume that we start from the interacting condensate with
$\lambda=\mathrm{const}(t)$, then at $t=0$ suddenly apply a $\pi$
phaseshift and follow the subsequent appearance of the ``cat''.
Clearly if $\lambda>\lambda_c$ (see~(\ref{lambdac})) the system
will become unstable and the quantum fluctuations will force it
into the entangled state. Without any calculations, it is obvious
that the final state will not be maximally entangled as described
in Fig.~\ref{cat:fig4} because of the energy conservation. Indeed,
the state with all the bosons occupying one well costs a huge
interaction energy which can not be compensated unless there is an
external pumping resulting in the time dependence of the
hamiltonian. From the same considerations it is obvious that the
maximum possible entanglement within this scheme can be achieved
at some intermediate values of $\lambda$. Thus if
$\lambda<\lambda_c$, there is no instability to begin with and so
no entanglement in the final state. On the other hand for the
large $\lambda$ any significant number fluctuations will cost a
strong increase in the interaction energy, which can not be
compensated by a limited decrease of the hopping energy. We plot
the resulting number variance as a function of time for the
periodic array of ten wells in Fig.~\ref{cat:fig5a}. The critical
value of the interaction for this case is
$\lambda_c=2\sin^2(\pi/10)\approx 0.19$. Clearly for $\lambda$
getting closer to $\lambda_c$ the oscillations become slower and
the steady state is achieved later. We would like to point out,
that this type of dynamics certainly corresponds to a sudden
perturbation discussed in Sec.~\ref{sec:num}. Therefore we expect
a finite ergodic time for a full quantum solution so that there is
no true steady state for the finite number of bosons. Therefore to
achieve a stable stationary entangled state it is always
preferable to drive the system adiabatically towards the
instability, i.e. apply a $\pi$-phaseshift at smallest possible
$\lambda$ and then slowly ramp up the interaction.

\subsubsection{Harmonic Trap}

The effect of harmonic trap can be mimicked by adding a quadratic
potential term to the hamiltonian of the system (\ref{hamilt}):
\be
V_j={J\omega_t^2\over 4}j^2.
\ee
Here $\omega_t$ is the trapping frequency.

As before let us assume that initially the interactions are
suppressed and the system is in the ground state:
\be
|O\rangle =\left(\sum_j \alpha_j a_j^\dagger\right)^{N_{tot}}
|Vac\rangle,
\ee
where $|Vac\rangle$ denotes a vacuum state with no particles,
$N_{tot}$ is the total number of bosons in the system and
\be
\alpha_j\approx {\sqrt{2}\over (\pi\omega_t)^{1\over 4}}\,\mathrm
e^{-{\omega_t^2\over 4}j^2}
\ee
is the ground state wave function of a single boson in the
harmonic trap in the coordinate representation. The state
$|O\rangle$ can be also written as:
\be
|O\rangle =\int {d\theta\over 2\pi}\mathrm e^{i\theta
(n_1+n_2+\dots -N_{tot})} |\sqrt{N_{tot}}\alpha_1\rangle_c |{\sqrt
N_{tot}}\alpha_2\rangle_c\dots,
\ee
where as usual $|\alpha\rangle_c$ stands for the coherent state.
If the number of populated wells is not small, then we can ignore
the global phase and use an approximate expression
\be
|O\rangle \approx \prod_j |\sqrt{N_{tot}}\alpha_j\rangle_c .
\ee
As in the previous discussion we assume that at initial time the
phases in the adjacent wells were uniformly shifted by some phase
$\phi$. However, we will not consider only the case with
$\phi=\pi$, because the $\pi$ phase shift in the nonuniform
potential does not give a stationary solution, although it still
preserves inversion symmetry. Also the number variance is no
longer a convenient measure of the instability, because the
distribution is not uniform even in the initial state. Instead let
us introduce two other quantities: coordinate of the center of
mass and the condensate width:
\be
X={1\over N}\langle \sum_j j a_j^\dagger a_j\rangle, \;
W=\sqrt{{1\over N} \langle \sum_j j^2 a_j^\dagger a_j\rangle-X^2}.
\label{eq50}
\ee
The semiclassical operators can be trivially obtained from
(\ref{eq50}). For example,
\be
X_{cl}={1\over N} \langle \sum_j j\,
(\psi_j^\star\psi_j-1/2)\rangle.
\ee
Let us assume that both the interaction and the trap frequency
increase in time:
\be
\lambda(t)={\tanh\delta t\over 1-\delta t},\quad
\omega_t^2(t)={\omega_t^2(0)\over 1 - \delta t}.
\label{eq52}
\ee
We define $\lambda$ according to (\ref{deflambda}) with
$N=N_{tot}|\alpha_0|^2$ being the number of bosons in a central
well. Note that with $\omega_t\neq 0$ there are two degrees of
freedom: $\lambda/J$ and $\omega_t/J$. If we want to reflect the
experimentally relevant case of vanishing tunnelling rather then
increasing interaction we have to keep the ratio
$\lambda/\omega_t^2$ constant. Extra multiplier ``$\tanh\delta
t$'' in the interaction term is taken only for convenience
purposes and it does not qualitatively change any of the results.

\begin{figure}
\centering
\includegraphics[width=8.3cm]{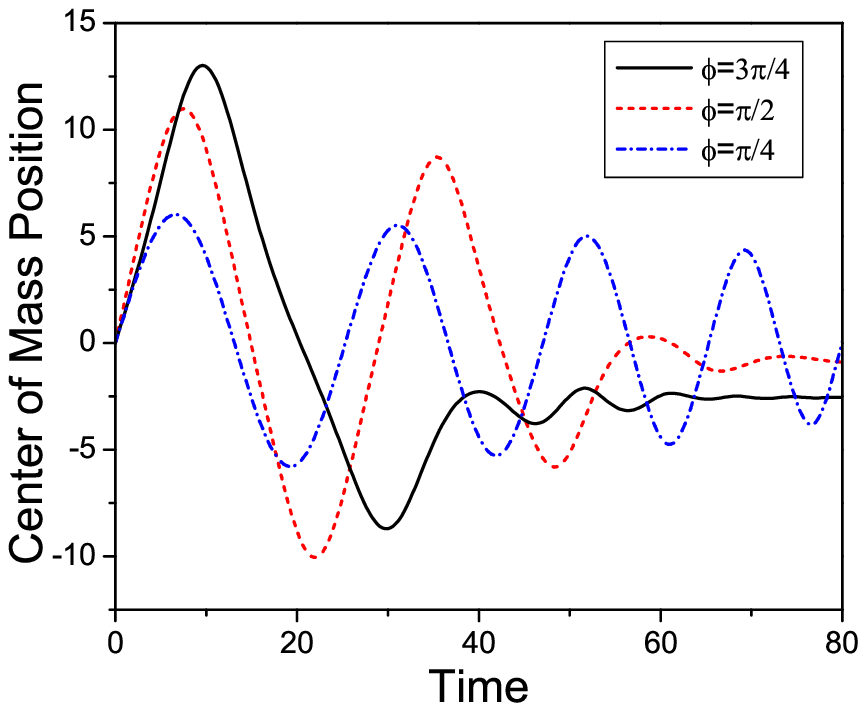}
\includegraphics[width=8.3cm]{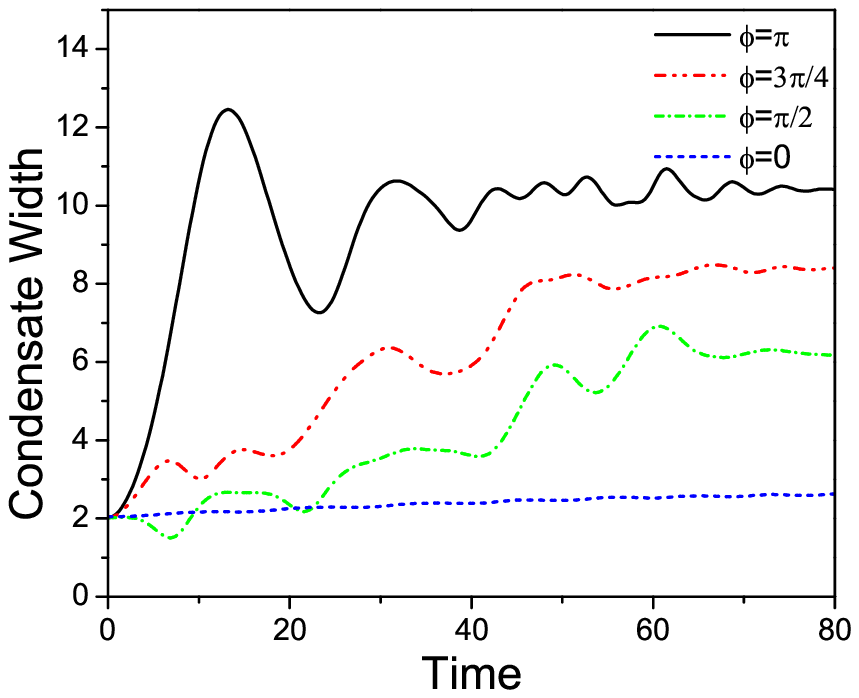}
\caption{Center of mass position and the condensate width as a
function of time for the interaction and trapping frequency
increasing in time according to (\ref{eq52}) with $\delta=0.01$,
$\omega_t^2(0)=0.03$, and the total number of bosons
$N_{tot}=100$.}
\label{cat:fig5}
\end{figure}

The two graphs in Fig.~\ref{cat:fig5} show the condensate width
and the center of mass position versus time for different initial
phase imprints. Note that if the phaseshift is equal to $\pi$,
there is no center of mass oscillation because of the inversion
symmetry. However the width oscillations are very large and
pronounced in this case. They also die very fast away from the
$\pi$ phaseshift, however the steady state condensate width
depends on $\phi$ rather smoothly. It is also interesting to note
that the center of mass oscillations decay faster for larger
angles. This is in the direct agreement with the previous
predictions of self-trapping at large
phase-gradients~\cite{Smerzi}. We would like to point out that the
trapping is real in our case because of the rapidly increasing
interaction (or equivalently vanishing tunnelling).

\begin{figure}
\centering
\includegraphics[width=8.3cm]{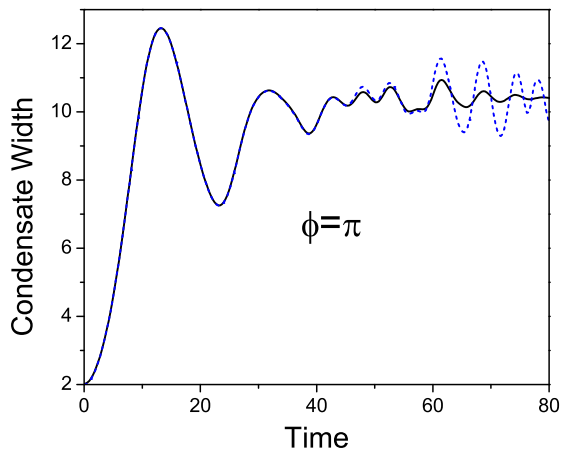}
\caption{Condensate width as a function of time for the same
conditions as in Fig.~(\ref{cat:fig5}). The dashed line is the
simple Gross-Pitaevskii result and the solid line is the improved
calculation according to (\ref{eq25}).}
\label{cat:fig6}
\end{figure}

There is, however, a major difference between the results for a
periodic array and a harmonic trap. In the classical limit the
$\pi$-state is an eigenstate in a periodic array for any strength
of the interaction, therefore the motion is completely driven by
the quantum-mechanical fluctuations, while in a harmonic trap the
$\pi$ state is classically unstable (that is why we plot our
observables versus interaction for the periodic array and versus
time for the parabolic trap). The classical instability makes
quantum effects unimportant if the number of bosons is large. To
further elucidate this point let us compare the results for a
single Gross-Pitaevskii trajectory, i.e. simple classical limit,
with the TWA result. The two dependences are shown in
Fig.~\ref{cat:fig6} and they are clearly very alike. This proves
that the resulting instability has a simple classical nature and
does not give a ``cat'' state, rather the condensate simply has
large amplitude breathing modes. Such an outcome should not be
very surprising. In order to get to the macroscopical entanglement
one has to prepare a classically equilibrium but unstable state,
which can be driven away via quantum fluctuations. This can be
achieved by a simple phase shift only in a uniform periodic
lattice.

\begin{figure}
\centering
\includegraphics[width=8.3cm]{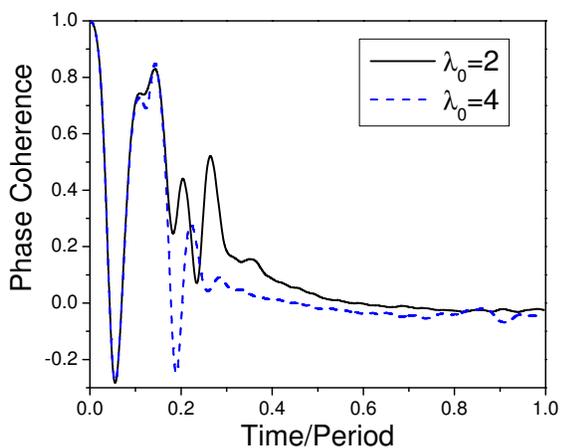}
\caption{Phase coherence as a function of time for the periodic
interaction~(\ref{addt}) with $\delta=0.004$. The total number of
bosons $N_{tot}=100$, the trap frequency $\omega_t^2=0.03$, and
the initial phaseshift $\phi=\pi$. Clearly the dynamics is
completely irreversible as opposed to Fig.~\ref{cat:fig4a}.}
\label{cat:fig7}
\end{figure}

We can also check the reversibility of the evolution for the
periodic in time interaction~(\ref{addt}). And clearly, (see
Fig.~\ref{cat:fig7}) the dynamics is completely irreversible. This
result is also not surprising given that the initial state is not
stationary even without interactions, so the $\pi$ phaseshift
looks as a sudden perturbation applied at $t=0$ and even if the
interactions change with time slowly the system is not in the
adiabatic regime.

\section{Discussion}

In the previous sections we adopted a reduced Wigner approximation
allowing to treat dynamical evolution of classical instabilities
due to quantum fluctuations. In this analysis we completely
ignored the classical noise, which may come from various sources
e.g. fluctuations of the laser intensity or wavelength,
fluctuations of the magnetic field creating the harmonic trap,
collisions with external atoms, triple interactions between the
bosons leading to the loss of the atoms from the condensate, etc.
All these sources are generally quite weak, otherwise no beautiful
experiments would ever be possible. On the other hand the
``cat''states are extremely fragile to any kinds of classical
noise. For example, we know that in a double well in the
equilibrium the noise required to destroy the coherence
exponentially scales with the mass (or equivalently number of
particles). So that heavy objects are always localized in one of
the classical minima. However, this is not the case in our
situation. Indeed, the system is far from the equilibrium and the
``cat''state is not a ground state. Let us crudely estimate the
limits on the noise. Assume that the classical fluctuations result
in a random external potential with a usual correlator $\langle
Y(t)Y(t^\prime)\rangle = Y_0 \delta (t-t^\prime)$ and consider the
case of two wells for simplicity. Then the relative phase flow
with time due to fluctuations will be
\begin{displaymath}
\delta\theta\sim N Y_0\sqrt{t}.
\end{displaymath}
On the other hand the minimum time required to get to the cat
state is (see Sec.~\ref{sec:2}):
\begin{displaymath}
t\sim {1\over \delta}\sim \ln N.
\end{displaymath}
Requiring that the total phase accumulated during the evolution is
less then one we get the upper bound for the noise which does not
destroy the coherence :
\begin{displaymath}
Y_0\sim {1\over N\sqrt{\ln N}}.
\end{displaymath}
This result is encouraging, because instead of exponentional we
get a much weaker scaling of the noise with the number of
particles. Clearly, in the multi-well case there will be an extra
scaling with the number of wells and of course no cat state is
possible in the thermodynamic limit. However, the scaling will be
again weak and tractable. Another possibility to observe
macroscopic ``cat'' states experimentally is to use a modulated
hopping between the sites in a large array effectively splitting
the condensate into pairs of sites, so the effects of the
classical noise become weaker.

The other two constraints used in our analysis that the number of
bosons is large and the condensate is one-dimensional are also not
essential. We used the large number of bosons rather to satisfy
the formalism then to explain the effect. The instability is there
even in the fully quantum-mechanical treatment of the problem and
the resulting maximally entangled state is clearly independent of
$N$. The only thing is that the evolution of macroscopically
entangled states with a large number of atoms is more interesting
from the experimental prospect. Concerning higher dimensions we
would like to point out that, although we did not performed actual
calculations, it is extremely unlikely that any of the results
will qualitatively change. The instability will clearly survive in
any dimensions and the $\pi$-state can be implemented, at least
theoretically, in square lattices of arbitrary dimensionality.

In conclusion, we showed that it is possible to create
macroscopically entangled ``Schr\"odinger cat'' states in the
bosonic systems in optical lattices with finite number of sites.
We justified and used truncated Wigner approximation generalizing
a simple GP approach by the exact treatment of quantum
fluctuations at initial time of the evolution. The resulting
expressions were tested on a solvable model of the two coupled
condensates and we found a very good agreement with the exact
results. At the end we presented some numerical simulations for
the multiple-well condensates and argued that it is possible to
create a ``cat'' state with more than two possible outcomes.

\begin{acknowledgments}
The author would like to acknowledge helpful discussions with E.
Altman, I. Carusotto, A. Clerk, S.~Girvin, M. Kasevich,
S.~Sachdev, K.~Sengupta, A. Tuchman. This research was supported
by US NSF Grant DMR 0098226.
\end{acknowledgments}

\appendix
\section{Microscopic derivation of the truncated Wigner approximation}
\label{twa}

Let us assume that at the initial time of evolution $t=0$ the
system is described by some density matrix $\rho_0$:
\be
\rho_0=\sum_\chi P(\chi)|\chi\rangle \langle \chi|,
\ee
where $|\chi\rangle$ represents some basis and $P(\chi)$ is the
probability to be in a particular state (if we use coherent basis
then $P(\chi)$ coincides with Glauber P-function widely used in
quantum optics~\cite{Walls}). If the initial state is pure, the
sum contains only single term. According to standard
quantum-mechanical formulas the expectation value of an arbitrary
normal-ordered operator $\Omega$ at arbitrary time $t$ is given
by:
\be
\Omega(t)=\mathrm{Tr}\left\{\rho_0 T_{K\tau}\mathrm e^{i \int_0^t
\mathcal{H}(\tau)d\tau} \Omega \mathrm e^{-i\int_0^t \mathcal{
H}(\tau)d\tau}\right\},
\label{eq21}
\ee
where $T_{K\tau}$ is the time-ordering operator along the Keldysh
contour going from $0$ to $t$ and then returning back to $0$,
$\mathcal H$ is the Hamiltonian of the system, e.g. (\ref{hamilt})
in our case. In the same way one can define correlation functions
of products of operators. The conventional trick to deal with
expressions like (\ref{eq21}) is to rewrite them in the
path-integral form using the coherent-state
representation~\cite{Negele}. The only difference with the more
usual equilibrium case is that there are two exponents containing
$\mathcal H$. So it is convenient to introduce two fields $a_f$
and $a_b$ propagating forward and backward in time~\cite{Kamenev}.
Then instead of (\ref{eq21}) we can write
\begin{widetext}
\beq
&&\Omega(t)=\int Da_f Da_b\, \langle a_{b\,0}|\rho_0 |
a_{f\,0}\rangle\, \mathrm e^{-a_{f\,0}^\star
a_{f\,0}+a_{f\,0}^\star a_{f\,1}+i\mathcal H
(a_{f\,0}^\star,a_{f\,1})\Delta\tau}\!\!\dots \mathrm
e^{-a_{f\,Q}^\star a_{f\,Q}}\nonumber\\
&&~~~\Omega(a_{f\,Q}^\star, a_{b\,Q},t)\mathrm e^{a_{f\,Q}^\star
a_{b\,Q}} \mathrm e^{-a_{b\,Q}^\star a_{b\,Q} +a_{b\,Q}^\star
a_{b\,Q-1}-i\mathcal H(a_{b\,Q}^\star a_{b\,Q-1})\Delta\tau}\!\!
\dots \mathrm e^{-a_{b\,0}^\star a_{b\,0}},
\label{last22}
\eeq
\end{widetext}
where $\Delta\tau=t/Q$ and $Q\to\infty$, $\Omega(a_f^\star,a_b,t)$
is the normal ordered operator $\Omega$ with the fields
$a^\dagger(t)$ and $a(t)$ substituted by the complex numbers
$a_f^\star(t)$ and $a_b(t)$. The expression above is intentionally
written in the discrete form, since we want to take special care
of the boundary effects. In the classical limit the evolution
equations are deterministic, therefore
$a_{f\,cl}(\tau)=a_{b\,cl}(\tau)$. On the other hand, because of
quantum fluctuations the two trajectories may be different.
Therefore instead of the forward and backward fields it is natural
to introduce their classical ($\psi$) and quantum ($\eta$)
combinations~\cite{Kamenev}: $a_f=\psi+{\eta/ 2},\;a_b=\psi-{\eta/
2}$. So that in the classical limit $\psi$ should correspond to
the solution of the GP equations and $\eta$ should be simply equal
to zero. Now we can take a continuum limit in (\ref{last22})
taking $\Delta\tau\to 0$ so that:
\begin{widetext}
\beq
&&\langle \Omega(t)\rangle \approx\int D\eta(\tau) D\psi(\tau)\,
\langle \psi_0-{\eta_0\over 2}|\rho_0 | \psi_0+{\eta_0\over
2}\rangle\, \Omega(\psi(t)^\star+{\eta(t)^\star\over
2},\psi(t)-{\eta(t)\over 2})\nonumber\\
&&~~~\mathrm e^{-|\psi_0|^2-{1\over 4}|\eta_0|^2-{1\over
2}|\eta(t)|^2}\mathrm e^{{1\over
2}(\eta_0^\star\psi_0-\eta_0\psi_0^\star)}\mathrm e^{\int_0^t
d\tau\, \eta^\star(\tau)\mathcal L[\psi,\psi^\star,\tau] -
\eta(\tau)\mathcal L^\star[\psi,\psi^\star,\tau]}\mathrm e^{
\int_0^t d\tau {U(\tau)\over 4}
\left(\psi^\star(\tau)\eta(t)+\psi(\tau)\eta^\star(\tau)\right)|\eta(\tau)|^2},
\label{last23}
\eeq
\end{widetext}
where $\mathcal L[\psi,\psi^\star,\tau]$ stands for the classical
(GP) differential operator acting on the field $\psi(t)$:
\be
\mathcal L_j[\psi,\psi^\star,\tau]\equiv i{d \psi_j\over d \tau}+
(\psi_{j+1}+\psi_{j-1})-\lambda(\tau)\, \psi_j^\star \psi_j^2.
\ee
Note that $\mathcal L$ as well as fields $\psi$ and $\eta$ contain
spatial indices which we suppressed in (\ref{last23}) to simplify
notations. A closer look to equation (\ref{last23}) shows that
there are linear, quadratic and cubic terms in the quantum field
$\eta$ the latter appearing only due to interactions. It is
intuitively clear that in the classical limit $\eta(\tau)$ should
be small in some sense. Thus if we ignore completely all nonlinear
terms in $\eta$, then the functional integral over the quantum
field becomes a trivial product of $\delta$-functions enforcing GP
equations on the classical field $\psi$. The next approximation
will be to leave quadratic corrections but ignore cubic.
From~(\ref{last23}) it is clear that the quadratic corrections
affect only the initial and the final times of the evolution, that
is why it was important for us to start from a discrete version of
the path integral and be careful about boundaries. The integral
over $\eta(t)$ transforms the operator $\Omega$ into $\Omega_{cl}$
according to (\ref{eq26c}). It is a simple exercise to check that
$\Omega_{cl}$ is obtained by first symmetrizing $\Omega$ and then
substituting the operators $a$ and $a^\dagger$ by the c-numbers
$\psi$ and $\psi^\star$. For example if
\be
\Omega=a^\dagger a= {1\over 2}(a^\dagger a+a a^\dagger)-{1\over 2}
\ee
then
\be
\Omega_{cl}=\langle \psi^\star+{\eta^\star/ 2},\psi-{\eta/
2}\rangle=\psi^\star\psi-{1\over 2},
\ee
where as it follows from (\ref{last23}), the average over $\eta$
is taken with the weight $\exp(-|\eta|^2/2)$.

The second quadratic contribution originates from the field
$\eta_0$ corresponding to the initial time of evolution. Because
of the coupling to ($\psi_0$) in (\ref{last23}), these
fluctuations introduce a probability distribution for the
classical initial conditions given by~(\ref{eq26a}). If we ignore
the corrections to the classical equations of motion coming from
the third power of the quantum field $\eta$, then the time
dependence of the observable $\Omega$ will be given by
(\ref{eq25}).

Let us now give general comments of validity of (\ref{eq25}) and
(\ref{eq26a}). If the Hamiltonian is non-interacting, then these
expressions are exact. If there are nonlinear interactions, then
in general, there will be corrections to the action involving
terms proportional to all odd powers of $\eta$: $\eta^3$ in our
case (see (\ref{last23})) or higher in general. Those terms will
affect the time evolution, which will not be described by the GP
equations any longer. In~Ref.[\onlinecite{ap1}] we show that the
corrections to the GP dynamics arise in the form of the quantum
scattering events, which are equivalent to the nonlinear response
to the infinitesimal perturbation of the field $\psi$ along its
classical path. We only note here that these corrections are
always of the form $f(t)/N$ with $f(t)$ being some time dependent
function satisfying $f(t)\to 0$ at $t\to 0$, and $1/N$ is our
semiclassical parameter. So the TWA given by~(\ref{eq25}) and
(\ref{eq26a}) always gives the exact short time asymptotical
behavior of the evolution. As we discuss in Sec.~\ref{subsec:two},
the time when TWA breaks down depends on the details of a
particular process and becomes longer under slow perturbations,
where only a limited number of quantum levels are excited.

Another point we would like to make is that though the expansion
in powers of $1/N$ is clearly around the classical solution, there
is no $\hbar$ present anywhere. This should not be surprising
since here and quite often in the atomic physics the Planck's
constant either completely absorbed into energies, which are
measured in Hz, or into time. In conventional units $\hbar^{-1}$
appears as the prefactor in the action justifying the saddle-point
or classical approximation. In the same way the number of bosons
per site $N$ appears as a prefactor in the exponent of
(\ref{last23}) after rescaling $\psi\to\sqrt{N}\psi$ and
$\eta\to\sqrt{N}\eta$. So in general any expansion in powers of
$\eta$ is in fact the expansion in powers of $\hbar$.

Let us finally spend a few words discussing the difference between
the present derivation and that of the conventional Keldysh
technique. The key point in our discussion is that we ascribed the
time dependence to the operator $\Omega$ itself, while leaving the
density matrix time independent. This allowed us to completely
separate initial quantum fluctuations, which entered in the form
of Wigner distribution of initial conditions to classical
trajectories (\ref{eq26a}), from the quantum dynamical effects
(which we consider in~Ref.[\onlinecite{ap1}]). On the other hand
in the Keldysh technique the density matrix acquires time
dependence and the initial density matrix is absorbed into the
quantum propagator~\cite{Kamenev}. While it is still possible to
derive GP equations in the saddle-point approximation, integrating
out quantum fields in the lowest order gives a complicated
self-interacting classical action~\cite{Kamenev}, which is hardly
possible to deal with except perturbatively or using stochastic
methods. This should not be surprising since any diagrammatic
technique uses a noninteracting Gaussian limit as a starting
point. Therefore to get just a classical GP dynamics in the
Keldysh technique, it is necessary to sum all diagrams with
classical vertices (three classical fields and one
quantum)~\cite{Kamenev}, which looks virtually impossible.


\begin{thebibliography}{99}

\bibitem{Kasevich} C.~Orzel, A.~K.~Tuchman, M.~L.~Fenselau, M.~Yasuda,
and M.~A.~Kasevich, Science {\bf 291}, 2386 (2001).

\bibitem{Bloch}  M.~Greiner, O.~Mandel, T.~Esslinger, T.~W.~H\"ansch,
and I.~Bloch, Nature, {\bf 415}, 39 (2002).

\bibitem{Bloch1}  M.~Greiner, O.~Mandel, T.~W.~H\"ansch,
and I.~Bloch, Nature, {\bf 419}, 51 (2002).

\bibitem{Fisher} M.~P.~A.~Fisher, P.~B.~Weichman, G.~Grinstein, D.~S.~Fisher,
Phys. Rev. B {\bf 40}, 546 (1989).

\bibitem{Leggett} A.J.~Leggett, Rev. Mod. Phys., {\bf 73}, 307
(2001).

\bibitem{Rey}  A.M.~Rey, K.~Burnett, R.~Roth, M.~Edwards, C.J.~Williams, and
C.W.~Clark, cond-mat/0210550.

\bibitem{Smerzi} A.~Smerzi, A.~Trombettoni, P.G.~Kevrekidis, and
A.R.~Bishop, Phys. Rev. Lett., {\bf 89}, 170402 (2002).

\bibitem{Wu} B.~Wu and Q.~Niu, Phys. Rev. A {\bf 64}, 061603, (2001).

\bibitem{Raghavan} S.~Raghavan, A.~Smerzi, S.~Fantoni, and
S.~R.~Shenoy, Phys. Rev. A{\bf 59}, 620 (1999).

\bibitem{Kanamoto} R.~Kanamoto, H.~Saito, and M.~Ueda, Phys. Rev. A, {\bf 67}, 013608
(2003).

\bibitem{Kavoulakis} G.M.~Kavoulakis, Phys. Rev. A {\bf 67}, 011601
(2003).

\bibitem{psg} A.~Polkovnikov, S.~Sachdev, and S.M.~Girvin,
Phys. Rev. A {\bf 66}, 053607 (2002).

\bibitem{Vardi} J.R.~Anglin and A.~Vardi, Phys. Rev. A {\bf 64}, 013605 (2001).

\bibitem{Franzosi} R.~Franzosi and V.~Penna, cond-mat/0205209.

\bibitem{Sachdev2} S.~Sachdev, K.~Sengupta, and S.M.~Girvin, Phys. Rev. B {\bf 66},
075128 (2002).

\bibitem{Kasevich1} M.~Kasevich, private communication.

\bibitem{Walls} D.F.~Walls and G.J.~Milburn, {\em Quantum Optics}, Springer-Verlag, Berlin (1994).

\bibitem{Steel} M.J.~Steel, M.K.~Olsen, L.I.~Plimak, P.D.~Drummond,
S.M.~Tan, M.J.~Collet, D.F.~Walls, and R.~Graham, Phys. Rev. B
{\bf 58}, 4824 (1998).

\bibitem{Sinatra} A.~Sinatra, Y.~Castin, and C.~Lobo, J. of Mod. Optics {\bf 47},
2629 (2000).

\bibitem{Sinatra1} A.~Sinatra, C.~Lobo, and Y.~Castin, Phys. Rev. Lett. {\bf 87},
210404  (2001).

\bibitem{Sinatra2} A.~Sinatra, C.~Lobo, and Y.~Castin, cond-mat/0201217.

\bibitem{Lobo} C.~Lobo, A.~Sinatra, and Y.~Castin, cond-mat/0301628.

\bibitem{Stoof1} H.T.C. Stoof, J. of Low Temp. Phys., {\bf 114},
11 (1999).

\bibitem{Duine}  R.A. Duine, H.T.C. Stoof, cond-mat/0211514.

\bibitem{Carusotto1} I.~Carusotto, Y.~Castin, and J.~Dalibard, Phys. Rev. A {\bf 63},
023606 (2003).

\bibitem{Carusotto2} I.~Carusotto and Y.~Castin, cond-mat/0209135.

\bibitem{ap1} A. Polkovnikov, cond-mat/0303628.

\bibitem{Kamenev} A.~Kamenev, cond-mat/0109316.

\bibitem{Caldeira} A.O. Caldeira, A.J. Leggett, Physica {\bf
121A}, 587 (1983).

\bibitem{Gardiner} C.W. Gardiner, {\em Quantum Noise},
Springer-Verlag, Berlin Heidelberg (1991).

\bibitem{Dorner}  U.~Dorner, P.~Fedichev, D.~Jaksch, M.~Lewenstein, and
P.~Zoller, quant-ph/0212039.

\bibitem{Negele} J.W.~Negele and H.~Orland,
{\em Quantum many-particle systems}, Addison-Wesley, Redwood City
(1988).

\end{thebibliography}
\end{document}